\documentclass[prd,aps,preprint,nofootinbib,superscriptaddress]{revtex4}
\usepackage{epsfig}
\usepackage{amsmath}
\input{epsf}

\begin{document}


\vspace*{2cm}
\title{A note on the scale evolution of tri-gluon correlations}

\author{Andreas Sch\"{a}fer}
\affiliation{\normalsize\it Institut f\"{u}r Theoretische Physik,Universit\"{a}t Regensburg, Regensburg, Germany}
\author{Jian Zhou}
\affiliation{\normalsize\it Institut f\"{u}r Theoretische Physik,Universit\"{a}t Regensburg, Regensburg, Germany}

\begin{abstract}
We reexamine the scale dependence  of the gluonic pole tri-gluon correlations
relevant for the transverse single spin asymmetries. Our results differ from those of  all previous
similar calculations.
The  corresponding asymptotic behavior of tri-gluon correlations at small $x$ is also briefly discussed.
\end{abstract}

\maketitle
Twist-3 quark-gluon and tri-gluon correlation functions play the important role for the theoretical description
of transverse single spin asymmetries (SSAs) in the framework of collinear factorization.
In order to  test QCD dynamics and reduce the dependence of theory predictions on the energy scale,
it is  necessary to calculate the scale evolution of these nonperturbative functions.
Another motivation is that numerical studies of the their scale evolution might help to build models for these
twist-3 correlations.
The scale evolution of twist-3 quark gluon correlations relevant for SSAs have been derived in
Refs.~\cite{Kang:2008ey,Zhou:2008mz,Vogelsang:2009pj,Braun:2009mi,Ma:2011nd}.
The result obtained in Ref.~\cite{Braun:2009mi}  differs from that derived in
Refs.~\cite{Kang:2008ey,Zhou:2008mz,Vogelsang:2009pj,Ma:2011nd}.
After recovering an additional boundary term obtained in~\cite{Braun:2009mi}
 by the different groups within their different approaches~\cite{Schafer:2012ra,Ma:2012xn,Kang:2012em,Ma:2012ye,Kang:2012ns},
 an agreement on the scale evolution of twist-3 quark-gluon correlation functions  was finally reached.

On the other hand, the scale dependence of tri-gluon correlations relevant for SSAs has also been studied in Refs.~\cite{Kang:2008ey,Braun:2009mi}.
We recently have shown that also in this case,  a term was missed~\cite{Schafer:2013wca}. This term depends on the hard gluon pole contributions from quark-gluon correlations.
In this short contribution, we focus on the flavor non-singlet case, for which our results do not agree with~\cite{Kang:2008ey,Braun:2009mi}.
Interestingly, our newly derived evolution equation takes a closed form at small x.
The corresponding asymptotic behavior of tri-gluon correlations at small x is  briefly discussed.

The tri-gluon correlations relevant for SSAs are defined through the following matrix element~\cite{Ji:1992eu,Kang:2008qh},
\begin{eqnarray}
 T_G^{(\pm)}(x,x)=g_0\int \frac{dy_1^- dy_2^-}{2\pi} e^{ixp^+y_1^-}
 \frac{1}{xp^+} g_{\alpha \beta}\epsilon_{S \gamma n p}
 \langle pS| C_{\pm}^{bca} F_b^{\beta+}(0)F_c^{\gamma +}(y_2)F_a^{\alpha+}(y_1^-)|pS \rangle \ ,
\end{eqnarray}
where $S$ is the nucleon transverse spin vector, normalized as $S^2=-1$, and $p^\mu, n^\mu$ are commonly defined lightlike vector.
All gluon polarization indices $\alpha, \beta$ and $\gamma$ are regarded to be transverse.
$C_+^{bca}$ and $C_-^{bca}$ are respectively the symmetric and anti-symmetric structure constants the color SU(3) group,
\begin{equation}
C_+^{bca}=if^{bca} , \ \ \ \ \ \ \ \ C_-^{bca}=d^{bca}  \ .
\end{equation}
At  order  $\alpha_s$, tri-gluon correlations contributing to the above gluonic pole matrix elements can be parameterized as follows
\cite{Ji:1992eu,Beppu:2010qn},
\begin{eqnarray}
&&-g_0i^3\int \frac{dy^- dz^-}{(2\pi)^2 p^+} e^{iy^- x_1p^+}e^{iz^-(x_2- x_1)p^+}
\langle pS| d^{bca} F_b^{\beta+}(0)F_c^{\gamma +}(z^-)F_a^{\alpha+}(y^-)|pS \rangle
 \nonumber \\
&=&2iM_N \left [ O(x_1,x_2)g^{\alpha \beta}\epsilon^{\gamma p n S}
+O(x_2,x_2-x_1)g^{\beta \gamma}\epsilon^{\alpha p n S}
+O(x_1,x_1-x_2)g^{\gamma \alpha} \epsilon^{\beta p n S} \right ] \ ;
\\
&&-g_0i^3\int \frac{dy^- dz^-}{(2\pi)^2 p^+} e^{iy^- x_1p^+}e^{iz^-(x_2- x_1)p^+}
\langle pS| i f^{bca} F_b^{\beta+}(0)F_c^{\gamma +}(z^-)F_a^{\alpha+}(y^-)|pS \rangle
 \nonumber \\
&=&2iM_N \left [ N(x_1,x_2)g^{\alpha \beta}\epsilon^{\gamma p n S}
-N(x_2,x_2-x_1)g^{\beta \gamma}\epsilon^{\alpha p n S}
-N(x_1,x_1-x_2)g^{\gamma \alpha} \epsilon^{\beta p n S} \right ] \ .
\end{eqnarray}
From this decomposition, it is easy to verify the relations~\cite{Beppu:2010qn}:
\begin{eqnarray}
 \frac{x}{2 \pi} T_G^{(+)}(x,x)&=&-4M_N \left [ N(x,x)- N(x,0) \right ]  \ ,
 \\
 \frac{x}{2 \pi} T_G^{(-)}(x,x)&=&-4M_N \left [ O(x,x)+ O(x,0) \right ] \ .
\end{eqnarray}
As shown below, it is more convenient to write the  scale evolution equations in terms of $ \left [ N(x,x)- N(x,0) \right ]$ and
$\left [ O(x,x)+ O(x,0) \right ]$ rather than $ T_G^{(+)}(x,x)$ and $ T_G^{(-)}(x,x)$.
We will first deduce evolution equations for $ \left [ N(x,x)- N(x,0) \right ]$ and
$\left [ O(x,x)+ O(x,0) \right ]$, and then  reexpress them in terms of $ T_G^{(+)}(x,x)$ and $ T_G^{(-)}(x,x)$ in the small x limit.

We carry out the calculation  in the light cone gauge with Mandelstam-Leibbrandt (ML) prescription~\cite{Mandelstam:1982cb,Leibbrandt:1983pj}.
The gluon propagator according to the ML prescription reads,
\begin{eqnarray}
D^{\alpha\beta}(l)=\frac{-i}{l^2+i\epsilon} \left [ g^{\alpha\beta} -
\frac{(l^\alpha n^\beta +n^\alpha l^\beta)(l \cdot p)}{(l \cdot n )(l \cdot p)+ i\epsilon}
\right ] \ .
\end{eqnarray}
The DGLAP evolution equations for unpolarized quark and gluon distributions have been reproduced in the light-cone gauge with the ML prescription~\cite{Bassetto:1996ph}.
We compute the splitting kernel for the  evolution of tri-gluon correlations by applying  the same technique as used in ~\cite{Bassetto:1996ph}.

\begin{figure}[t]
\begin{center}
\includegraphics[width=16cm]{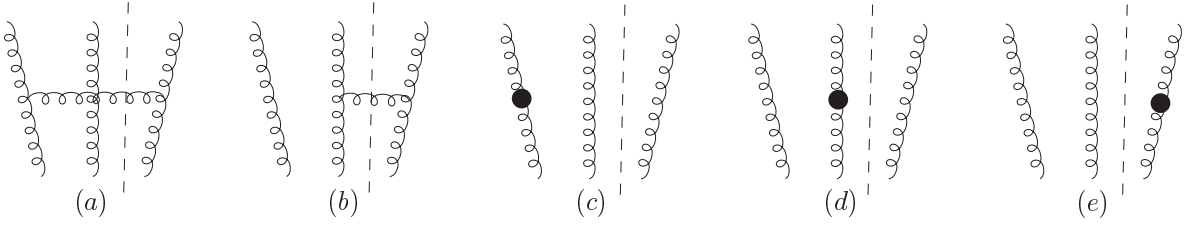}
\end{center}
\vskip -0.4cm \caption{Feynman diagrams contributing to the scale
evolution of  gluonic pole tri-gluon correlations in the light cone
gauge with Mandelstam-Leibbrandt prescription. Black dots represent
a gluon or quark loop.} \label{fig1}
\end{figure}

All Feynman diagrams contributing to the scale evolution of  gluonic pole tri-gluon correlations in the light-cone
gauge with Mandelstam-Leibbrandt prescription are illustrated in the Fig.1. Fig.1(a) and Fig.1(b) give rise to real radiation  while
Fig.1(c), Fig.1(d) and Fig.1(e) are the gluon self energy  diagrams. The contribution from all other virtual diagrams vanish because multiple
poles lie in the same half plane~\cite{Ma:2011nd}.

The calculation of Fig.1(a) is rather straightforward. Its contribution to the splitting kernels is given by,
\begin{eqnarray}
\frac{\partial \left [ N(x,x)- N(x,0) \right ]}{\partial {\rm ln} \mu_F^2} |_{\rm Fig.1(a)}
   =
\frac{\alpha_s}{2\pi} \int_x^1 \frac{d \xi}{\xi}
C_A  \frac{(z^2-z+1)^2}{(1-z)_+} \left [ N(\xi,\xi)
 -  N(\xi,0)\right ] \ ,
\end{eqnarray}
and,
\begin{eqnarray}
\frac{\partial \left [ O(x,x)+ O(x,0) \right ]}{\partial {\rm ln} \mu_F^2} |_{\rm Fig.1(a)}
=
\frac{\alpha_s}{2\pi} \int_x^1 \frac{d \xi}{\xi}
 C_A  \frac{(z^2-z+1)^2}{(1-z)_+}\left [ O(\xi,\xi) +   O(\xi,0)\right ]  \ ,
\end{eqnarray}
where $z=x/\xi$, and $\mu_F$ is the factorization scale. The plus prescription is defined in the usual way.
Fig.1(b) generates the hard gluon pole contributions,
\begin{eqnarray}
&& \!\!\!\!\!\!\!\!\!\!\!\!\!\!\!\!\!\!
\frac{\partial \left [ N(x,x)- N(x,0) \right ]}{\partial {\rm ln} \mu_F^2} |_{\rm Fig.1(b)} =
\frac{\alpha_s}{2\pi} \int_x^1 \frac{d \xi}{\xi}
 \frac{C_A}{2} { \bigg \{ }  -2\delta(1-z)\left [N(x,x)-N(x,0)\right ]
 \nonumber \\ &&
  +\frac{1+z^2}{(1-z)_+} N(x,\xi)
  - \frac{1+(1-z)^2}{(1-z)_+} N(\xi,\xi-x)
 -\frac{z^2+(1-z)^2}{(1-z)_+}N(x,x-\xi) { \bigg \} }.
\end{eqnarray}
For O-type functions, one has,
\begin{eqnarray}
&&  \!\!\!\!\!\!\!\!\!\!\!\!\!\!\!\!\!\!
\frac{\partial \left [ O(x,x)+ O(x,0) \right ]}{\partial {\rm ln} \mu_F^2} |_{\rm Fig.1(b)} =
\frac{\alpha_s}{2\pi} \int_x^1 \frac{d \xi}{\xi} \frac{C_A }{2}
\left \{ -2\delta(1-z)\left [O(x,x)+O(x,0)\right ]  \right .\
 \nonumber \\ && \left .\
 + \frac{1+z^2}{(1-z)_+} O(x,\xi)+ \frac{1+(1-z)^2}{(1-z)_+} O(\xi,\xi-x)
 +\frac{z^2+(1-z)^2}{(1-z)_+}O(x,x-\xi) \right \}   .
\end{eqnarray}
Here we would like to mention that one has to perform the calculation with great care in order not to loose
the boundary terms proportional to $\delta(1-z)$ which appear in the above two equations.
We refer  interested readers to Refs.~\cite{Braun:2009mi,Schafer:2012ra}
 for the techniqcal details how to derive these terms in the light-cone gauge with a certain prescription.

Fig.1(c), Fig.1(d) and Fig.1(e) give rise to the virtual contributions.  The one loop gluon self energy correction
has been evaluated in the light cone gauge with ML prescription in~\cite{Dalbosco:1986eb}. It is easy to extract the collinear divergence
from the complete result.  We then obtain,
\begin{eqnarray}
\frac{\partial \left [ N(x,x)- N(x,0) \right ]}{\partial {\rm ln} \mu_F^2} |_{\rm Fig.1(c)+Fig.1(d)+Fig.1(e)} &=&
\frac{\alpha_s}{2\pi}
\left ( C_A \frac{11}{6}-\frac{n_f}{3} \right ) \left [N(x,x)-N(x,0)\right ],
\\
\frac{\partial \left [ O(x,x)+ O(x,0) \right ]}{\partial {\rm ln} \mu_F^2} |_{\rm  Fig.1(c)+Fig.1(d)+Fig.1(e)} &=&
\frac{\alpha_s}{2\pi} \left ( C_A \frac{11}{6}-\frac{n_f}{3} \right )
\left [O(x,x)+O(x,0)\right ] .
\end{eqnarray}
where $n_f$ is the number of active quark flavors. Note that in the above equations,
 the coupling constant $g_0$ appearing in the matrix element definition of tri-gluon correlations
is replaced by the renormalized one,  $g$, satisfying the equation
$\partial g/\partial {\rm ln} \mu_F^2  = -g(\alpha_s/4\pi) [C_A 11/6-n_f/3]  $.

Summing up all contributions, we obtain the scale evolution equations for $\left [ N(x,x)- N(x,0) \right ]$ and $\left [ O(x,x)+ O(x,0) \right ]$.
Now let's turn to discuss the small $x$ behavior of tri-gluon correlations. At leading logarithmic ${\rm ln} x$ accuracy, one may make the following
approximations:
\begin{eqnarray}
&& N(x,\xi) \approx N(0, \xi) , \ \ N(\xi,\xi-x) \approx N(\xi, \xi),  \ \ N(x,x-\xi)\approx N(0,-\xi) ; \\
&& O(x,\xi) \approx O(0, \xi) , \ \ O(\xi,\xi-x) \approx O(\xi, \xi),  \ \ O(x,x-\xi)\approx O(0,-\xi) .
\end{eqnarray}
With these approximations and using the symmetry relations~\cite{Beppu:2010qn},
\begin{eqnarray}
&& N(x_1,x_2) = N(x_2, x_1) , \ \ N(x_1,x_2) = -N(-x_1,-x_2),  \\
&&  O(x_1,x_2) = O(x_2, x_1) , \ \ O(x_1,x_2) = O(-x_1,-x_2),
\end{eqnarray}
the evolution equations are  dramatically simplified in the small $x$ limit,
\begin{eqnarray}
\frac{\partial \left [ N(x,x)- N(x,0) \right ]}{\partial {\rm ln} \mu_F^2}  &=&
\frac{\alpha_s}{2\pi} \int_x^1 \frac{d \xi}{\xi} C_A (-z) \left [ N(\xi,\xi)- N(\xi,0) \right ],
\\
\frac{\partial \left [ O(x,x)+ O(x,0) \right ]}{\partial {\rm ln} \mu_F^2} &=&
\frac{\alpha_s}{2\pi} \int_x^1 \frac{d \xi}{\xi} C_A 2
\left [O(\xi,\xi)+O(\xi,0)\right ] .
\end{eqnarray}
It turns out that both equations take a closed form at small $x$.
These evolution equations also can be reexpressed in terms of the tri-gluon functions $T_G^{(+)}(x,x)$ and $T_G^{(-)}(x,x)$,
\begin{eqnarray}
\frac{\partial T_G^{(+)}(x,x)}{\partial {\rm ln} \mu_F^2}  &=&
\frac{\alpha_s}{2\pi} \int_x^1 \frac{d \xi}{\xi} C_A  (-1) T_G^{(+)}(\xi,\xi),
\\
\frac{\partial T_G^{(-)}(x,x)}{\partial {\rm ln} \mu_F^2} &=&
\frac{\alpha_s}{2\pi} \int_x^1 \frac{d \xi}{\xi} C_A \frac{2}{z}
T_G^{(-)}(\xi,\xi).
\end{eqnarray}
Based on  these results, we make the following predictions:

i) The tri-gluon correlation function $T_G^{(+)}(x,x)$ can be related to the $k_\perp$ moment of the gluon Sivers function. Since the splitting kernel is negative,
one might expect that the gluon Sivers function changes sign \footnote{We in particular thank Vladimir
Braun, Alexander Manashov and Bjorn Pirnay for illuminating discussions on this point. } at small $x$.

ii) The function $xT_G^{(+)}(x,x)$ will drop rapidly with decreasing $x$ due to the complete cancelation of leading pole terms proportional to $1/z$.

iii) One notices that the splitting kernels for tri-gluon correlation $T_G^{(-)}(x,x)$ and the unpolarized gluon distribution $G(x)$ are identical at small $x$.
As a consequence, we anticipate that within the DGLAP formalism and up to leading order accuracy, the function $T_G^{(-)}(x,x)$ rises with decreasing $x$ in the same manner as
 $G(x)$.

To summarize: We reexamined the scale evolution of tri-gluon correlations relevant for transverse single spin asymmetries.
Our results differ from those of all previous calculations, i.e., quite disappointingly, the agreement reached
at last for the scale evolution of twist-3 quark-gluon correlations does not yet extend  to the tri-gluon correlations case
and more work will be needed to settle this issue.

Based on the results presented in this short note, we find that spin asymmetries are dominated by the tri-gluon function $T_G^{(-)}(x,x)$, or more generally by the function $O(x_1,x_2)$, at small $x$.
The small $x$ asymptotic behavior of  $T_G^{(-)}(x,x)$
is expected to be the same as for the unpolarized gluon distribution within the DGLAP formalism.
Furthermore, as byproduct of this work, it has been shown that the gluon Sivers  function could change sign at small $x$.
These predictions can be tested by measuring  spin dependent asymmetries
 in various channels in ep and pp collisions at RHIC
and a future EIC~\cite{Kang:2008qh,Kang:2008ih,Beppu:2010qn,Koike:2011ns,Yuan:2008vn,Hatta:2013wsa}.

\noindent
{\bf Acknowledgments:} We are grateful to Vladimir
Braun and Alexander Manashov for valuable discussions on the various relevant topics.
J. Zhou also thanks Bjorn Pirnay for his help during the preparation of this manuscript.
This work has been supported by BMBF (05P12WRFTE).

\end {document}